\documentstyle[12pt]{article}
\title{Siegel superparticle, higher order\\
fermionic constraints, and path integrals}
\author{Anton V. Galajinsky\thanks{galajin@fma.if.usp.br}  and 
Dmitri M. Gitman\thanks{gitman@fma.if.usp.br}\\
Instituto de Fisica, Universidade de S\~ao Paulo,\\
C. Postal 66318, 05315-970, S\~ao Paulo, SP, Brasil}
\date{}
\begin{document}
\maketitle
\large
\begin{abstract}
We study Siegel superparticle moving in $R^{4|4}$ flat superspace. 
Canonical quantization is accomplished yielding the massless 
Wess-Zumino model as an effective field theory. Path integral 
representation for the corresponding superpropagator is constructed 
and proven to involve the Siegel action in a gauge fixed form. It is 
shown that higher order fermionic constraints intrinsic in the theory,
though being a consequence of others in $d=4$, make a crucial
contribution into the path integral.
\end{abstract}
\vspace{0.2cm}

\noindent



\vspace{1cm}

\section{Introduction}

During the past decade a great deal of interest has been attracted
to the investigation of different aspects of quantization of
relativistic particles. The extensive researches were mainly focused
in two directions. The first line began with the works [1,2] where a 
pseudoclassical model (PM) for the Dirac particle in four dimensions
was constructed, studied, and quantized. The chief goal was to see how
the quantum mechanics of a spin--$\frac 12$ particle, which may also be
treated as a free spinor field theory, can be reconstructed in the
course of first quantization of Grassmann classical mechanics
(pseudoclassical mechanics). A number of papers generalizing the
results to various spins and divers dimensions appeared [3--7], which can 
generally be 
regarded as answering the questions: Is it possible to treat any field
theory (at least a free field theory) as the result of first
quantization of a classical model or a PM? Does there exist a regular
method to construct such a model for a given field theory? 
In the relation to the latter point it is worth mentioning that the 
construction of a path integral representation for a propagator of a
field theory may serve as a heuristic method to find such a 
PM. The effectiveness of this approach was
demonstrated for the scalar and spinor particles [8] and then used to 
construct a new minimal PM for spinning particles in odd dimensions
[7] (for the related works on path integral quantization of the models
see also Refs. 9,10).

Another line of research originated from the superstring
theory where the problem of consistent covariant quantization proved
to have its analog in more simple mechanics examples. Of prime interest 
(for motivations and a comprehensive list of references see Ref. 11)
seems to be the Casalbuoni-Brink--Schwarz (CBS) superparticle [12] which 
involves an infinitive ghost tower when $BRST$ quantized. 
As is known, the puzzle of covariant quantization of the model
can be addressed in two respects [11] (for an alternative twistor-harmonic
approach see Ref. 13): The problem of quantizing the infinitely
reducible first--class constraints; The problem of quantizing the infinitely
reducible second--class ones. It is the Siegel
superparticle [14], later referred to as $AB$--superparticle, which
allows the studying of the former in an independent way. Since the 
$AB$--formulation involves only first--class constraints of the CBS
theory, the two models are not equivalent [15]. However, in Ref. 16 further 
modifications including higher order fermionic constraints 
($ABC$--, $ABCD$--superparticles) have been
proposed and shown [17] to be equivalent to the CBS model. As
the new formulations possess first--class constraints only, the 
second--class constraints problem intrinsic in the CBS theory can be
avoided.

Quantization of the $10d$ $ABCD$--superparticle both by $BV$ and $BFV$
methods has been accomplished in a series of work [18--20]. In particular,
cohomologies of the $BRST$ operator have been evaluated [19] giving the
ten-dimensional super Yang-Mills multiplet in the result. It is to be
mentioned, however, that in both approaches the expression for the 
effective action involves an infinite tower of ghost variables and,
hence, looks formal.

In the present paper we study the $AB$--superparticle in $R^{4|4}$ flat
superspace. As shown below, this dimension is unique in the sense that
the analog of the higher order fermionic $C$--constraint of the $10d$
$ABC--$,$ABCD--$formulations (which removes negative norm states from 
quantum spectrum [16,17]) automatically holds in the $4d$ $AB$--model.
Although it is not of direct relevance to superstring theory (since 
beyond of critical dimension), quantization of the theory suggests
an instructive example of generalizing the results of the theory of 
spinning particles to supersymmetric area. By now, only a few articles
treating a precise relation between (world volume) supersymmetric 
mechanics and corresponding field theory superpropagators are known 
[21,22]. 

There is one more motivation for this work. In a recent paper [23] a 
recipe how to supplement infinitely reducible first class constraints
up to a constraint system of finite stage of reducibility has been 
proposed. It may suggest an efficient way to cure the
infinite ghost tower problem intrinsic in the $BRST$ quantized Siegel 
theory. We expect that the knowledge of propagators of the first
quantized theory will provide a considerable simplification in the
forthcoming path integral test of this approach.

The work is organized as follows. In Sec. 2 Hamiltonian analysis of
the model is presented. Complete constraint system is 
found showing that the analog of the $C$--constraint of the $10d$
$ABC$--, $ABCD$--formulations is a consequence of others in $d=4$.
Light-cone quantization of the model is examined in Sec. 3. Quantum
spectrum proves to contain for essentially different states, the
corresponding helicities being $(-\frac 12,0,0,\frac 12)$. In Sec. 4
in the course of Dirac quantization we reproduce the results of Sec. 3
in a covariant fashion. An effective field theory corresponding to
the first quantized $4d$ Siegel superparticle proves to be the 
massless Wess-Zumino model in the component form. Structure of the
superfield representation realized in the quantum theory is discussed 
in Sec. 5. Quantum states of the first quantized 
Siegel superparticle are shown to form a reducible representation of 
the super Poincar\'e group which contains superhelicities 0 and $-1/2$.
Path integral representation for propagator of the
first quantized theory is constructed in Sec. 6 and proven to involve
the Siegel action in a gauge fixed form. It is shown that higher order
fermionic constraints, though being a consequence of others in $d=4$, 
make a crucial contribution into the path integral. 
\section{Canonical formalism}
As originally formulated in the first order formalism the $4d$
Siegel superparticle action is [14] 
\begin{equation}
S=\int d\tau \{ p_m(\dot x^m+i\theta\sigma^m\dot{\bar\theta}-
i\dot\theta\sigma^m\bar\theta+i\psi\sigma^m\bar\rho-i\rho\sigma^m\bar\psi)-
\frac {ep^2}{2}-{\rho^\alpha}{\dot\theta}_\alpha-{\bar\rho}_{\dot\alpha}
{\dot{\bar\theta}}^{\dot\alpha} \},
\end{equation}
and can be put into Lagrangian form by removing $p^n$ with the use of
its equation of motion
\begin{equation}
S=\int d\tau \{ \frac 1{2e}(\dot x^m+i\theta\sigma^m\dot{\bar\theta}-
i\dot\theta\sigma^m\bar\theta+i\psi\sigma^m\bar\rho-i\rho\sigma^m\bar\psi)^2-
\rho^\alpha{\dot\theta}_\alpha-{\bar\rho}_{\dot\alpha}
{\dot{\bar\theta}}^{\dot\alpha} \}.
\end{equation}
The model is invariant under rigid supersymmetry transformations in
the standard realization
\begin{equation}
\delta\theta=\epsilon, \qquad \delta\bar\theta=\bar\epsilon, \qquad
\delta x^n=i\theta\sigma^n\bar\epsilon-i\epsilon\sigma^n\bar\theta,
\end{equation}
and with respect to local reparametrizations and $\kappa$-symmetry,
$$
\begin{array}{lll} \delta_\alpha\theta=\alpha\dot\theta, &
\delta_\alpha\bar\theta=\alpha\dot{\bar\theta}, &
\delta_\alpha x^n=\alpha\dot x^n,\\
\delta_\alpha\rho=\alpha\dot\rho, &
\delta_\alpha\bar\rho=\alpha\dot{\bar\rho}, &
\delta_\alpha e=(\alpha e)^\cdot,\\
\delta_\alpha\psi=(\alpha\psi)^\cdot, &
\delta_\alpha\bar\psi=(\alpha\bar\psi)^\cdot;\end{array}
\eqno{(4a)}$$
$$
\begin{array}{l}
\delta_\kappa\theta=\displaystyle -ie^{-1} \Pi_n\sigma^n\bar\kappa,
\qquad \delta_\kappa\bar\theta= ie^{-1} \Pi_n\kappa\sigma^n,\\
\delta_\kappa x^n=i\delta\theta\sigma^n\bar\theta
-i\theta\sigma^n\delta\bar\theta-i\kappa\sigma^n\bar\rho+
i\rho\sigma^n\bar\kappa,\\
\delta_\kappa e=4\dot\theta\kappa+4\bar\kappa\dot{\bar\theta}, \qquad
\delta_\kappa\psi=\dot\kappa,\\
\delta_\kappa\bar\psi=\dot{\bar\kappa},\end{array}
\eqno{(4b)}$$
\addtocounter{equation}{1}
where we denoted $\Pi^m=\dot x^m+i\theta\sigma^m\dot{\bar\theta}-
i\dot\theta\sigma^m\bar\theta+i\psi\sigma^m\bar\rho-i\rho\sigma^m\bar\psi$.

The meaning of the variables entering into the action (2) is quite
different and deserves to be mentioned here. The coordinates 
$(x^m,\theta^\alpha,\bar\theta_{\dot\alpha})$ parametrize the 
standard ${R}^{4|4}$ superspace. The variables $e$ and $(\psi^\alpha,
\bar\psi_{\dot\alpha})$ prove to be gauge fields for the local
$\alpha$-- and $\kappa$--symmetries respectively. The pair
$(\rho^\alpha, \bar\rho_{\dot\alpha})$ provides the terms
corresponding to (mixed) covariant propagator for fermions and, as
shown below, allows one to accomplish covariant quantization without 
lose of Lorentz covariance.

Passing to the Hamiltonian formalism one finds primary constraints in
the problem\footnote{We define momenta conjugate to fermi variables
to be right derivatives of the Lagrangian with respect to  
velocities. This corresponds to the following choice of the Poisson
bracket $\{\theta^\alpha,p_{\theta\beta}\}={\delta^\alpha}_\beta$,
$\{\bar\theta_{\dot\alpha},{p_{\bar\theta}}^{\dot\beta}\}=
{\delta_{\dot\alpha}}^{\dot\beta}$ and the position of momenta and
velocities in the Hamiltonian as specified below in Eq. (6).}
\addtocounter{equation}{1}
$$
\begin{array}{ll} {p_\theta}^\alpha-i{(p_n\sigma^n\bar\theta)}^\alpha-
{\rho}^\alpha=0, & p_{\rho\alpha}=0,\\
p_{\bar\theta\dot\alpha}+i{(\theta\sigma^np_n)}_{\dot\alpha}-{\bar\rho}_
{\dot\alpha}=0, & {p_{\bar\rho}}^{\dot\alpha}=0,
\end{array}
\eqno{(5a)}$$
$$
p_{e}=0, \qquad p_{\psi\alpha}=0, \qquad {p_{\bar\psi}}^{\dot\alpha}=0,
\eqno{(5b)}$$
where $(p,p_\theta,p_{\bar\theta},p_\psi,p_{\bar\psi},p_\rho,
p_{\bar\rho},p_e)$ are momenta canonically 
conjugate to the variables 
$(x,\theta,\bar\theta, \psi,\bar\psi,\rho,\bar\rho,e)$ respectively.
The total Hamiltonian has the form
\begin{eqnarray}
%\lefteqn{
&& H^{(1)}=p_e\lambda_e+p_{\psi\alpha}{\lambda_\psi}^\alpha
+{p_{\bar\psi}}^{\dot\alpha}\lambda_{\bar\psi\dot\alpha}
+p_{\rho\alpha}{\lambda_\rho}^\alpha+{p_{\bar\rho}}^{\dot\alpha}
\lambda_{\bar\rho\dot\alpha}\cr
&& +{(p_{\bar\theta}+i\theta\sigma^np_n-\bar\rho)}^{\dot\alpha}
\lambda_{\bar\theta\dot\alpha}+
{(p_\theta-ip_n\sigma^n\bar\theta-\rho)}_\alpha{\lambda_\theta}^\alpha\cr
&& +e\displaystyle\frac{p^2}2-i\psi\sigma^n\bar\rho p_n+
i\rho\sigma^n\bar\psi p_n,
\end{eqnarray}
where $\lambda_{\dots}$ denote Lagrange multipliers corresponding to the
primary constraints. The consistency conditions for the primary constraints
imply the secondary ones
\begin{equation}
p^2=0, \qquad p_n\sigma^n\bar\rho=0, \qquad \rho\sigma^np_n=0,
\end{equation} and determine some of the Lagrange multipliers,
\begin{equation}
\begin{array}{ll} \lambda_\theta=-ip_n\sigma^n\bar\psi,& \qquad
\lambda_{\bar\theta}=i\psi\sigma^np_n,\\
\lambda_\rho=-2ip_n\sigma^n\lambda_{\bar\theta}\approx0,& \qquad
\lambda_{\bar\rho}=2i\lambda_\theta\sigma^np_n\approx0.
\end{array}
\end{equation}
No tertiary constraints arise at the next stage of the Dirac
procedure, the remaining Lagrange multipliers being unfixed. 

Thus, the complete constraint set of the model is given by Eqs. (5),(7) and 
it is convenient to rewrite the latter in the equivalent form
\begin{equation}
p^2=0, \qquad p_m\tilde\sigma^mp_\theta=0, \qquad
p_m\sigma^m p_{\bar\theta} =0.
\end{equation}
The constraints $(5a)$ are second--class and allow one to omit the pairs
$(\rho,p_\rho)$, $(\bar\rho,p_{\bar\rho})$ after
introducing the associated Dirac bracket
\begin{eqnarray}
\lefteqn{\{A,B\}_D=\{A,B\}-\{A,p_\rho\}\{p_\theta-ip_n\sigma^n
\bar\theta-\rho,B\}}\cr
&& -2i\{A,p_\rho\}\sigma^np_n\{p_{\bar\rho},B\}+\{A,p_\theta-
ip_n\sigma^n\bar\theta-\rho\}\{p_\rho,B\}\cr
&& -2i\{A,p_{\bar\rho}\}\sigma^np_n\{p_\rho,B\}-\{A,p_{\bar\rho}\}
\{p_{\bar\theta}+i\theta\sigma^np_n-\bar\rho,B\}\cr
&& +\{A,p_{\bar\theta}+i\theta\sigma^np_n-\bar\rho\}\{p_{\bar\rho},B\}.
\end{eqnarray}
The Dirac brackets for the remaining variables prove to coincide with the 
Poisson ones. The first--class constraints $(5b)$ admit the covariant gauge
\begin{equation}
e =1, \qquad \psi=0, \qquad \bar\psi=0,
\end{equation}
which yields
\begin{equation}
\lambda_e=0, \qquad \lambda_\psi=0, \qquad \lambda_{\bar\psi}=0,
\end{equation}
and implies that the canonical pairs $(e,p_e)$, $(\psi,p_\psi)$,
$(\bar\psi,p_{\bar\psi})$ can be eliminated from the consideration. 

Thus, after the partial
phase space reduction, there remain only $(x,p)$,
$(\theta,p_\theta)$, $(\bar\theta,p_{\bar\theta})$ variables 
being subject to the first--class constraints (9). The total Hamiltonian
vanishes on the constraint surface in the full agreement with the
reparametrization invariance of the model.

Some remarks are in order. First, by making use of a shift of the variables 
$\rho \rightarrow {\rho+i\sigma^n\bar\theta p_n}$, ${\bar\rho}\to
{\bar\rho-i\theta\sigma^n p_n}$, $e \to e+2(\psi\theta+\bar\theta\bar\psi)$  
the original Lagrangian (1) can be simplified to  
\begin{equation}
S=\int d\tau \{ p_m(\dot x^m+i\psi\sigma^m\bar\rho-i\rho\sigma^m\bar\psi)-
\frac {ep^2}{2}-{\rho^\alpha}{\dot\theta}_\alpha-{\bar\rho}_{\dot\alpha}
{\dot{\bar\theta}}^{\dot\alpha} \},
\end{equation}
or eliminating $p_n$
\begin{equation}
S=\int d\tau \{\frac 1{2e}(\dot
x^m+i\psi\sigma^m\bar\rho-i\rho\sigma^m\bar\psi)^2
-\rho^\alpha{\dot\theta}_\alpha-{\bar\rho}_{\dot\alpha}
{\dot{\bar\theta}}^{\dot\alpha} \}.
\end{equation}
In contrast to the theory (2), the formulation (14) proves to possess
a more simple and, in particular, off-shell closed algebra of 
local symmetries\footnote{The transformation with a local parameter 
$b^n$ is trivial on-shell.
It is the only trivial symmetry needed to close the algebra. The
closure of the algebra (4) is known to require an infinite number of
trivial symmetries.} (see also Ref. 24) 
$$
\begin{array}{l}
\delta_\kappa\theta=\displaystyle -ie^{-1} \Pi_n\sigma^n\bar\kappa,
\qquad \delta_\kappa\bar\theta= ie^{-1} \Pi_n\kappa\sigma^n,\\
\delta_\kappa x^n=i\rho\sigma^n\bar\kappa-i\kappa\sigma^n\bar\rho, \qquad
\delta_\kappa\psi=\dot\kappa,\\
\delta_\kappa\bar\psi=\dot{\bar\kappa},
\end{array}
\eqno{(15a)}$$
$$
\begin{array}{l}
\delta_b\theta=\displaystyle i b_n\sigma^n\dot{\bar\rho}, \qquad
\delta_b\bar\theta=-i b_n\dot\rho\sigma^n,
\end{array}
\eqno{(15b)}
$$
\addtocounter{equation}{1}
where $\Pi^m=\dot x^m+i\psi\sigma^m\bar\rho-i\rho\sigma^m\bar\psi$.
The rigid (on-shell) supersymmetry is realized in the theory in a 
nonstandard way
\begin{eqnarray}
&&\delta\theta=\epsilon, \qquad \delta\bar\theta=\bar\epsilon, 
\qquad 
\delta x^n=i\epsilon\sigma^n\bar\theta-i\theta\sigma^n\bar\epsilon \cr
&& \delta\rho=-ie^{-1} \Pi_n\sigma^n\bar\epsilon,\qquad 
\delta{\bar\rho}= ie^{-1} \Pi_n\epsilon\sigma^n, \cr
&& \delta e=-2(\psi\epsilon+\bar\epsilon\bar\psi),
\end{eqnarray}
and can be closed off-shell by applying the standard technique
[25]. It is sufficient to introduce an auxiliary vector variable $D^n$ which
transforms as $\delta_\kappa D^n=i(\dot\theta+i
e^{-1}(\Pi_n+D_n)\sigma^n\bar\psi)\sigma^m\bar\epsilon-
i\epsilon\sigma^m(\dot{\bar\theta}-i e^{-1}(\Pi_n+D_n)\psi\sigma^n)$, 
exchange $\Pi^m$ with $\Pi^m+D^m$ in Eq. (16) and add the trivial term
$-\frac 1{2e} D^2$ to the action (14).
As was shown in Ref. 24, the model (14) admits an interesting geometric
formulation that appeals to new extensions of the Poincar\'e superalgebra
in $d=3,4,6,10$ (in this relation see also Ref. 20). Beautiful
enough, it is this form of the Siegel action which appears in the path
integral when constructing a path integral representation for
superpropagator of the first quantized theory (see Sec. 6).

Second, from Eq. (9) it follows
\begin{equation} p_\theta^2=0, \qquad
p_{\bar\theta}^2=0.
\end{equation}
This means that the $C$--constraint of the
$10d$ $ABC$--,$ABCD$--superparticles, which removes negative norm 
states from quantum spectrum, automatically holds in four dimensions. In
particular, one can check that the model
\begin{eqnarray}
&& S=\int d\tau \{
p_m(\dot x^m+i\psi\sigma^m\bar\rho-i\rho\sigma^m\bar\psi)-\frac e2 p^2-
\rho\dot\theta-\bar\rho\dot{\bar\theta}\cr
&& +\lambda(\sigma+\tilde\sigma)
{(\rho-i\sigma^n\bar\theta p_n)}^2{(\bar\rho+i\theta\sigma^n p_n)}^2 \},
\end{eqnarray}
with $\lambda,\sigma, \tilde\sigma$ the new fermionic variables, is physically
equivalent to the original theory (1). It seems surprising, but the higher
order fermionic constraints (17), although being a consequence of
others in $d=4$, play an important role in quantum theory and make a
crucial contribution into the path integral.

\section{Ligh-cone quantization}

A covariant gauge to
Eq. (9) is known to be problematic in the original phase space.
Before turning to covariant quantization, it is worth 
considering the problem in the light-cone framework. 
This analysis
proves to suggest a correct choice of a superfield wave function when
accomplishing covariant quantization in the next section.
 
On the constraint surface $p^2=0$ only half of the fermionic constraints
(9) are linearly independent. Assuming the standard light-cone
condition $p^+\ne 0$, one finds them to be
\begin{equation}
p_{\theta2}- \frac{p^1+ip^2}{\sqrt{2}p^+} p_{\theta1}=0, \qquad
{p_{\bar\theta}}^{\dot1}+ \frac{p^1-ip^2}{\sqrt{2}p^+} 
{p_{\bar\theta}}^{\dot2}=0. 
\end{equation} 
After imposing the conventional light-cone gauge in the sector of 
fermi variables\footnote{We refrain from imposing a gauge to the 
bosonic first class constraint $p^2=0$ in order to maintain the
explicit connection between the Pauli-Lubanski vector and the momentum
vector in Eq. (30) below.} 
\addtocounter{equation}{1}
$$
\theta\sigma^+=0, \qquad  \sigma^+\bar\theta=0,\\
\eqno{(20a)}$$
or
$$
\theta^2=0, \qquad {\bar\theta}_{\dot1}=0,
\eqno{(20b)}$$
there remain only $(x^n,p_n)$,
$({\theta^1},p_{\theta1})$,$({\bar\theta}_{\dot2},{p_\theta}^{\dot2})$ 
variables\footnote{In what follows we omit the indices carried by 
the fermi variables.} subject to usual canonical commutation 
relations, provided the Dirac bracket associated with Eqs. (19),(20b) 
has been introduced. The gauge
fixed action is
\begin{equation}
S=\int d\tau \{ p_m\dot x^m+p_\theta \dot\theta+p_{\bar\theta}
\dot{\bar\theta}-\frac 12 p^2 \}.
\end{equation}

In passing to quantum description, a representation space {\it F}
for the fermi operators may be chosen to be a linear span\footnote{More
precisely, the {\it F} can be endowed with the structure of a
supervector space [26].} of four vectors
\begin{equation}
\left|0\right\rangle, \qquad \left|p\right\rangle, \qquad 
\left|\bar p\right\rangle,
\qquad \left|p,\bar p\right\rangle,
\end{equation}
with the action of the operators being defined like that of creation and
annihilation operators  
\begin{eqnarray}
&&
\hat\theta\left|0\right\rangle=0, \qquad \hat\theta\left|p\right\rangle=
i\left|0\right\rangle, \qquad \hat\theta\left|\bar p\right\rangle=0,
\qquad \hat\theta\left|p,\bar p\right\rangle=i\left|\bar p\right\rangle,\cr
&&
\hat{\bar\theta}\left|0\right\rangle=0, \qquad 
\hat{\bar\theta}\left|p\right\rangle=0,
\qquad \hat{\bar\theta}\left|\bar p\right\rangle=i\left|0\right\rangle,
\qquad \hat{\bar\theta}\left|p,\bar p\right\rangle=-i\left|p\right\rangle,\cr
&&
\hat {p_\theta}\left|0\right\rangle=\left|p\right\rangle, \qquad 
\hat {p_\theta}\left|p\right\rangle=0, \qquad
\hat {p_\theta}\left|\bar p\right\rangle=\left|p,\bar p\right\rangle,\qquad
\hat {p_\theta}\left|p,\bar p\right\rangle=0, \cr
&&
\hat {p_{\bar\theta}}\left|0\right\rangle=\left|\bar p\right\rangle, \qquad 
\hat {p_{\bar\theta}}\left|p\right\rangle=-\left|p,\bar p\right\rangle, \qquad
\hat {p_{\bar\theta}}\left|\bar p\right\rangle=0, \qquad    
\hat {p_{\bar\theta}}\left|p,\bar p\right\rangle=0.\nonumber \\
\end{eqnarray}
The total Hilbert space is defined to be a tensor product of the {\it F}
and the space of squire integrable functions in which $\hat{x^n}$ and
$\hat{p_n}$ act in the usual coordinate representation. An arbitrary
state is
\begin{equation}
\left|\psi\right\rangle=\left(\left|0\right\rangle a+
\left|p\right\rangle b+\left|\bar p\right\rangle c+\left|p,\bar
p\right\rangle d\right)\otimes\Phi(x),
\end{equation}
with $a,b,c,d$ the supernumbers. The conventional scalar product reads
[10]
\begin{equation}
\langle\psi_1|\psi_2\rangle=\int {d^4}x
\bar{\Phi_1}(x)\Phi_2(x)\left(\bar{a_1}a_2+\bar{b_1}b_2+\bar{c_1}c_2
+\bar{d_1}d_2\right),
\end{equation}
and can only formally be regarded as positive definite since involves
Grassmann numbers. Note that under this product 
${\hat {p_\theta}}^+=-i\hat\theta$, ${\hat {p_{\bar\theta}}}^+=
-i\hat{\bar\theta}$ which is in a agreement with the definition of
the operators like creation and annihilation ones.  

The physical subspace in the complete Hilbert space is defined by
the standard prescription [27,28]
\begin{equation}
\hat p^2|{\rm phys}\rangle=0.
\end{equation}

Thus, the Hilbert space of the model contains four essentially
different states. To evaluate helicities of the states we
reduce the Pauli-Lubanski vector
\begin{equation}
W_a=\frac12\epsilon_{abcd}p^bS^{cd}, \\
\end{equation}
were
$S^{cd}=p_{\theta\gamma}{\left(\sigma^{cd}\right)^\gamma}
_\delta\theta^\delta+{p_{\bar\theta}}^{\dot\gamma}
{{\left({\tilde\sigma}^{cd}\right)}_{\dot\gamma}}^{\dot\delta}{\bar\theta}
_{\dot\delta}$ is the spin part of the Lorentz generators, to the
surface of constraints (9) and gauges (20b). Making use of the identities 
\begin{equation}
\sigma_{ab}=\frac i2 \epsilon_{abcd}\sigma^{cd}, \qquad
{\tilde\sigma}_{ab}=-\frac i2 \epsilon_{abcd}{\tilde\sigma}^{cd},
\end{equation}
one gets
\begin{equation}
{W_a}=-\frac i2 p_a\left(p_{\theta}\theta-p_{\bar\theta}\bar\theta\right). 
\end{equation}
Passing further to quantum description, it is easy to verify that the 
relations 
\begin{eqnarray}
&&
\hat {W_a}\left|0\right\rangle\otimes\Phi(x)=
0\hat {p_a}\left|0\right\rangle\otimes\Phi(x), \cr
&& 
\hat {W_a}\left|p\right\rangle\otimes\Phi(x)=\frac 12 \hat {p_a}
\left|p\right\rangle\otimes\Phi(x), \cr
&&
\hat {W_a}\left|\bar p\right\rangle\otimes\Phi(x)=-\frac 12 \hat {p_a}
\left|\bar p\right\rangle\otimes\Phi(x), \cr
&&
\hat {W_a}\left|p,\bar p\right\rangle\otimes\Phi(x)=0 \hat {p_a}
\left|p,\bar p\right\rangle\otimes\Phi(x),
\end{eqnarray}
hold, provided the $pq$-ordering procedure for the
fermi operators has been adopted. Since $\hat{p^2}=0$ on physical
states, the equations above determine two particles of helecity 0 and 
two particles of helicities $\frac 12, -\frac 12$ respectively.
 
In the next section we consider Dirac quantization of the model. It is
the analysis above which supports our choice of a wave function to be a
{\it real scalar} superfield.

\section{Dirac quantization}

Since commutation relations for the variables $(x^n,p_m)$,
$(\theta^\alpha, p_{\theta_\alpha})$,
$(\bar\theta_{\dot\alpha}, {p_{\bar\theta}}^{\dot\alpha})$ are
canonical, we can
realize them in the coordinate representation
$\hat x^n=x^n$, $\hat p_m=-i\partial_m$,
$\hat\theta^\alpha=\theta^\alpha$,
$\hat p_{\theta_\alpha}=i\partial_{\alpha}$,
$\hat{{\bar\theta}_{\dot\alpha}}={\bar\theta}_{\dot\alpha}$,
$\hat {p_{\bar\theta}}^{\dot\alpha}=i\bar\partial^{\dot\alpha}$
on a Hilbert space whose elements are chosen to be real scalar superfields
\begin{eqnarray}
\lefteqn{V(x,\theta,\bar\theta)=A(x)+\theta\psi(x)+
\bar\theta\bar\psi(x)+\theta^2F(x)+\bar\theta^2
\bar F(x)}\cr
&& +\theta\sigma^n\bar\theta C_n(x)+\bar\theta^2\theta\lambda(x)
+\theta^2\bar\theta\bar\lambda(x)+
\theta^2\bar\theta^2D(x).
\end{eqnarray}
In what follows we assume the standard boundary condition
\begin{equation}
V(x,\theta,\bar\theta)\mathop{\longrightarrow}\limits_{x\to\pm\infty}0\;.
\end{equation}
The physical states in a complete Hilbert space are defined by Dirac's
prescription
\begin{eqnarray} && \hat p^2|{\rm phys}\rangle=0,\cr &&
\tilde\sigma^n\hat p_n\hat p_\theta|{\rm phys}\rangle=0,\\ &&
\tilde\sigma^n\hat p_n\hat p_{\bar\theta}|{\rm phys}\rangle=0.\nonumber
\end{eqnarray}
In the representation chosen this yields
$$
\begin{array}{l} \tilde\sigma^n\partial_n\psi=0, \qquad
\tilde\sigma^n\partial_n\bar\psi=0,\\
\Box A=0;\end{array}
\eqno{(34a)}$$
$$
\begin{array}{l}
(\tilde\sigma^m\sigma^n)^{\dot\alpha}{}_{\dot\beta}\partial_m C_n=0,
\qquad (\sigma^n\tilde\sigma^m)_\alpha{}^\beta\partial_m C_n=0,\\
\Box C_n=0,
\end{array}
\eqno{(34b)}$$
\addtocounter{equation}{1}
with all other component fields vanishing due to the boundary condition
(32). In obtaining Eq. (34) we used the identity
\begin{equation}
{\rm Tr}\,(\sigma^n\tilde\sigma^m)=-2\eta^{nm}.
\end{equation}
It is instructive then to simplify Eq. $(34b)$. Taking a trace in the
first equation and making use of Eq. (35) one finds \begin{equation}
\partial^nC_n=0,
\end{equation}
which (with the use of the relation $\sigma^n\tilde\sigma^m+
\sigma^m\tilde\sigma^n=-2\eta^{nm}$) brings Eq. $(34b)$ to the form
\begin{equation}
\begin{array}{l}
(\sigma^{mn})_\alpha{}^\beta\partial_m C_n=0, \qquad
(\tilde\sigma^{mn})^{\dot\alpha}{}_{\dot\beta}\partial_m C_n=0,\\
\Box C_n=0\end{array}
\end{equation}
Multiplying the first equality in Eq. (37) by
$(\sigma^{kl})_\beta{}^\alpha$ and taking into account the identity
\begin{equation}
{\rm Tr}\,\sigma^{mn}\sigma^{kl}=-\frac 12(\eta^{mk}\eta^{nl}-
\eta^{ml}\eta^{nk})-\frac i2\epsilon^{mnkl},
\end{equation}
one gets
\begin{equation}
\partial_kC_l-\partial_lC_k=-i\epsilon_{klmn}\partial^mC^n.
\end{equation}
Together with its complex conjugate this implies
\begin{equation}
\partial_mC_n-\partial_nC_m=0, \qquad \epsilon_{klmn}\partial^mC^n=0.
\end{equation}
The only solution to Eqs. (36), (40) is
\begin{equation}
C_n=\partial_nB,
\end{equation}
with $B$ the on-shell massless real scalar field
\begin{equation}
\Box B=0.
\end{equation}
Thus, physical states of the first quantized Siegel superparticle look
like
\begin{equation} V_{\rm
phys}(x,\theta,\bar\theta)=A(x)+\theta\psi(x)
+\bar\theta\bar\psi(x)+\theta\sigma^n\bar\theta
\partial_nB(x),
\end{equation}
with $A,B$ the on-shell massless real scalar fields (irreps of the
Poincar\'e group of helicity 0) and $\psi,\bar\psi$ the on-shell massless
spinor fields (helicities 1/2 and --1/2, respectively). Note that together
they fit to form two irreducible representations of the super Poincar\'e
group of superhelicities 0 and --1/2 [26].

It is worth mentioning that Eq. (33) can be rewritten in the manifestly
superinvariant form
\begin{equation}
\tilde\sigma^{n\dot\alpha\alpha}\partial_nD_\alpha V=0, \qquad
\tilde\sigma^{n\dot\alpha\alpha}\partial_n\bar D_{\dot\alpha}V=0,
\end{equation}
where $D_\alpha$, $\bar D_{\dot\alpha}$ are the covariant derivatives,
or as a single massless Dirac equation
\begin{equation}
\gamma^n\partial_n\Psi=0,
\end{equation}
with $\Psi\equiv\left(\begin{array}{c} D_\alpha V\\
\bar D^{\dot\alpha}V\end{array}\right)$ a (superfield) Majorana spinor.

An effective field theory which reproduces equations $(34a)$, (42) is
easy to write
\begin{equation}
S=\int d^4x\Big\{\frac 12 \partial^mA\partial_mA+ \frac 12 \partial^mB
\partial_mB+i\psi\sigma^m\partial_m\bar\psi\Big\},
\end{equation}
which is invariant under global (on-shell) supersymmetry
transformations
\begin{eqnarray}
&& \delta A=\epsilon\psi+\bar\epsilon\bar\psi, \qquad
\delta B=i\epsilon\psi-i\bar\epsilon\bar\psi,\cr
&& \delta\psi=i(\sigma^n\bar\epsilon)\partial_nA+
(\sigma^n\bar\epsilon)\partial_nB,\\
&& \delta\bar\psi=-i(\epsilon\sigma^n)\partial_nA+
(\epsilon\sigma^n)\partial_nB.\nonumber
\end{eqnarray}
In Eq. (46) we recognize the massless Wess--Zumino model in the
component form [29].

\section{Reducibility of the superfield representation and superhelicities}
As is known, the superfield formulation of the massless Wess--Zumino
model involves chiral and antichiral superfields [30],
$$ S = \int d^8z\, \Phi\bar\Phi,
\eqno{(48a)}$$
$$ \bar D_{\dot\alpha}\Phi=0,
\eqno{(48b)}$$
$$ D_\alpha\bar\Phi=0.
\eqno{(48c)}$$
The equations of motion read
$$ D^2\Phi=0,
\eqno{(49a)}$$
$$ \bar D^2\bar\Phi=0.
\eqno{(49b)}$$
\addtocounter{equation}{2}
Let us  show that the real scalar superfield (31) satisfying the
constraints (44) is the sum of on-shell massless chiral ($(48b)$,$(49a)$)
and antichiral ($(48c)$,$(49b)$) superfields.

Let us consider Eqs. $(48b), (49a)$. The first of them implies the
decomposition [26,30]
\begin{equation}
\Phi(x,\theta,\bar\theta)=\alpha(x)+\theta\psi(x)+\theta^2f(x)+
i\theta\sigma^n\bar\theta\partial_n\alpha(x)+\frac 12 \theta^2
\bar\theta\tilde\sigma^n\partial_n\psi+\frac 14\theta^2
\bar\theta^2\Box\alpha,
\end{equation}
while the latter, being rewritten in the equivalent form
\begin{equation}
\tilde\sigma^{n\,\dot\alpha\alpha}\partial_nD_\alpha\Phi=0,
\end{equation}
yields
\begin{equation}
\tilde\sigma^n\partial_n\psi=0, \qquad \Box\alpha=0, \qquad f=0.
\end{equation}
In order to get Eqs. (51), (52) we used the identity
\begin{equation}
[D^2,\bar D_{\dot\alpha}]=-4i{\sigma^n}_{\alpha\dot\alpha}
\partial_nD^\alpha,
\end{equation}
and assumed the standard boundary condition. Note also that Eqs.
$(48b)$, (51) together with the identity $\{ D_\alpha,\bar
D_{\dot\alpha}\}=-2i\sigma^n{}_{\alpha\dot\alpha} \partial_n$ imply
\begin{equation}
\Box\Phi=0,
\end{equation}

Thus, an on-shell chiral superfield can be written as
\begin{equation}
\begin{array}{c}
\Phi(x,\theta,\bar\theta)=\alpha(x)+\theta\psi(x)+i\theta\sigma^n\bar\theta
\partial_n\alpha(x),\\
\Box\alpha(x)=0, \qquad \tilde\sigma^n\partial_n\psi(x)=0.\end{array}
\end{equation}
Similarly, an on-shell antichiral superfield ($(48c)$,$(49b)$) has the form
\begin{equation}
\begin{array}{c}
\bar\Phi(x,\theta,\bar\theta)=\bar\alpha(x)+\bar\theta\bar\psi(x)-
i\theta\sigma^n\bar\theta\partial_n\bar\alpha(x),\\
\Box\bar\alpha(x)=0, \qquad \tilde\sigma^n\partial_n\bar\psi(x)=0.\end{array}
\end{equation}
Considering now the sum
\begin{equation}
\Phi+\bar\Phi=(\alpha+\bar\alpha)+\theta\psi+\bar\theta\bar\psi+
\theta\sigma^n\bar\theta\partial_ni(\alpha-\bar\alpha),
\end{equation}
and denoting
\begin{equation}
\alpha+\bar\alpha\equiv A, \qquad i(\alpha-\bar\alpha)\equiv B,
\end{equation}
one arrives just at Eq. (43).

Thus, the real scalar superfield subject to the constraints (44) 
is the sum of on-shell chiral and antichiral superfields
\begin{equation}
V_{\rm phys}(x,\theta,\bar\theta)=\Phi(x,\theta,\bar\theta)+
\bar\Phi(x,\theta,\bar\theta).
\end{equation}
This is in complete agreement with the results of the previous
section. It is worth mentioning that the choice of a wave function
to be a {\it complex scalar} superfield would not reproduce the 
results (see also the related work [31]).

As is known, on-shell massless scalar chiral superfields form a massless
irreducible representation of the super Poincar\'e group of
superhelicity $0$ [26]. Analogously, on-shell massless scalar antichiral
superfields realize irrep of superhelicity $-1/2$. We finally conclude
that quantum states of the first quantized
Siegel superparticle form a reducible representation of the super
Poincar\'e group which contains superhelicities 0 and $-1/2$.

Two remarks are relevant here. First, the constraints (9) 
just coincide with the first-class ones of the CBS superparticle. 
In Ref. 31 they have been used to covariantly 
quantize the CBS model within the framework of the Gupta-Bleuler 
method (see also the related work [32]). We are to stress, however, 
that the naive omitting of second--class constraints intrinsic in the 
CBS theory (which generally leads to the Siegel model) in the 
approach of Ref. 31 would not reproduce the result of the Dirac 
quantization presented above.  It is worth mentioning also Ref. 33, 
where the technique of quantization with a complex Hamiltonian has
been applied to establish a precise relation between on-shell massive 
chiral superfields and the corresponding particle mechanics. The massless
limit of the procedure, however, leads to ghost excitations in the
quantum spectrum [33] and, hence, is ill defined. \linebreak
Second, as was mentioned above the $C$--constraint
of the $10d$ $ABC$--,$ABCD$--superparticles is not necessary
in four dimensions. Note in this connection that an alternative
possibility 
$(p_\theta-ip_n\sigma^n\bar\theta)_{\alpha}
(p_{\bar\theta}+i\theta\sigma^np_n)_{\dot\alpha}=0$, or
$(D_\alpha\bar D_{\dot\alpha}-\bar D_{\dot\alpha}D_\alpha)V=0$
at the quantum level, leads to the trivial solution $V=0$ only (see,
however, Ref. [34]). By this reason, it is not obvious to us how to
extend the model (1) up to a theory equivalent to the $4d$ 
CBS superparticle along the lines of Ref. 16.

\section{Path integral representation of the superfield propagator}

Let examine now the possibility to reproduce the Siegel action within the
framework of a path integral representation for a (super)field propagator.
For the case concerned the superfield propagator reads [26,30]
\begin{eqnarray}
& G(z,z')=\left(\begin{array}{cc} 0 & G_c(z,z')\\
G_a(z,z') & 0\end{array}\right),\\
& G_c(z,z')=\displaystyle\frac 14\frac{\bar D^2}\Box \delta_-(z,z'),
\qquad \delta_-(z,z')=-\frac 14 D^2\delta^8(z-z'),\cr
& G_a(z,z')=\displaystyle\frac 14\frac{D^2}\Box \delta_+(z,z'),
\qquad \delta_+(z,z')=-\frac 14 \bar D^2\delta^8(z-z').\nonumber
\end{eqnarray}
Following Ref. 35 we represent it as a matrix element
\begin{equation}
G(z_{out},z_{in})=\langle z_{out}|\hat G |z_{in}\rangle,
\end{equation}
were $|z\rangle$ are eigenvectors of some coordinate operators
$\hat z^M=(\hat x^n,\hat\theta^\alpha,\hat {\bar\theta_{\dot\alpha}})$.
Together with the conjugate momenta
$\hat {p_M}=(\hat {p_n},\hat{p_{\theta\alpha}},
{\hat{p_{\bar\theta}}^{\dot\alpha}})$ they
satisfy the relations
\begin{eqnarray}
&&
\left[ \hat x^n,\hat p_m \right ]=i{\delta^n}_m, \qquad 
\{ \hat \theta^\alpha,
\hat p_{\theta\beta}\}=i{\delta^\alpha}_\beta, \qquad
\{ \hat {\bar\theta}_{\dot\alpha},
\hat {p_{\bar\theta}}^{\dot\beta} \}=i{\delta_{\dot\alpha}}^{\dot\beta},\cr
&&
\hat z^M \left|z\right\rangle=z^M \left|z\right\rangle, \qquad
\langle z|z' \rangle=\delta^8 (z-z'), \qquad \int d^8 z |z \rangle 
\langle z|=1, \cr 
&&
\hat p_M \left|p\right\rangle=p_M \left|p\right\rangle, \qquad
\langle p|p' \rangle=\delta^8 (p-p'), \qquad \int d^8 p |p \rangle 
\langle p|=1, \cr 
&&
-i\partial_m \langle z|\psi\rangle=\langle z|\hat p_m|\psi\rangle, \qquad
\qquad i\partial_\alpha \langle z|\psi\rangle=
\langle z|\hat p_{\theta\alpha}|\psi\rangle, \cr
&&
i\partial^{\dot\alpha} \langle z|\psi\rangle=
\langle z|{\hat{p_{\bar\theta}}}^{\dot\alpha}|\psi\rangle, \qquad
\qquad \langle z|\hat z^M|\psi\rangle=z^M \langle z|\psi\rangle, \cr
&&
\langle z|p \rangle=
\frac 1{\pi^2} e^{i p_n x^n+i p_{\theta\alpha}
\theta^\alpha+i
{p_{\bar\theta}}^{\dot\alpha}{\bar\theta}_{\dot\alpha}}\equiv
\frac 1{\pi^2} e^{i pz},
\end{eqnarray}
with $|\psi \rangle$ an arbitrary state. Derivatives with respect to
the odd variables entering into Eq. (62) are defined to be left ones. 
The simplest realization of the relations (62) is given by
$|\psi\rangle=\psi(z)=\psi(x,\theta,\bar\theta)$,
${|z\rangle}_{z_0}=\linebreak
=\delta^8 (z-z_0)$, $\langle\psi_1|\psi_2\rangle=
\int {d^8}z \bar{\psi_1}(z)\psi_2(z)$. In the explicit
representation (62) Eq. (61) acquires the form
\begin{eqnarray}
& G(z_{out},z_{in})=\frac 1{16} \langle z_{out}| 
\left(\begin{array}{cc} 0 & {({\hat p}^2)}^{-1} {\hat{p'_{\bar\theta}}}^2 
{\hat{p'_{\theta}}}^2 \\
{({\hat p}^2)}^{-1} {\hat{p'_{\theta}}}^2 {\hat{p'_{\bar\theta}}}^2 & 0
\end{array} \nonumber \right) |z_{in}\rangle,\\
\end{eqnarray}
where we denoted $\hat {p'_{\theta}}=\hat {p_{\theta}}-
i\hat{p_n} \sigma^n \hat{\bar\theta}$, $\hat {p'_{\bar\theta}}=
\hat {p_{\bar\theta}}+i\hat\theta \sigma^n \hat {p_n}$. Decomposing
then the matrix involved into the superpropagator in the sum of $1+1$
$\gamma$-matrices
\begin{eqnarray}
&&
\qquad \qquad \gamma^0=\left(\begin{array}{cc} 0 & 1\\
1 & 0\end{array}\right),\qquad
\gamma^1=\left(\begin{array}{cc} 0 & -1\\
1 & 0\end{array} \nonumber \right), \qquad \cr
&& 
\qquad \qquad \{ \gamma^n,\gamma^m \}=2\eta^{nm},
\qquad\eta^{nm}=(+,-); \\
&&
\left(\begin{array}{cc} 0 & {\hat{p'_{\bar\theta}}}^2 
{\hat{p'_{\theta}}}^2 \\
{\hat{p'_{\theta}}}^2 {\hat{p'_{\bar\theta}}}^2 & 0
\end{array} \nonumber \right)=
{\hat{p'_{\bar\theta}}}^2 {\hat{p'_{\theta}}}^2 \frac 1{\sqrt 2}\gamma^- +
{\hat{p'_{\theta}}}^2 {\hat{p'_{\bar\theta}}}^2 \frac 1{\sqrt 2}
\gamma^+, \\
\end{eqnarray}
where as usual $\gamma^\pm=\frac 1{\sqrt 2} (\gamma^0\pm\gamma^1)$,
we can conveniently rewrite Eq. (63) in the following equivalent form
\begin{equation}
G(z_{out},z_{in})=\frac 1{16 \sqrt 2} 
e^{\gamma^+ \frac \partial{\partial\mu}} e^{\gamma^-\frac 
\partial{\partial\nu}} \langle z_{out}|{({\hat p}^2)}^{-1} 
\left(\mu {\hat{p'_{\theta}}}^2 {\hat{p'_{\bar\theta}}}^2+      
\nu {\hat{p'_{\bar\theta}}}^2 {\hat{p'_{\theta}}}^2 \right) 
|z_{in}\rangle |_{\mu=\nu=0},
\end{equation}
where $\mu,\nu$ is a pair of auxiliary Grassmann
variables, the corresponding derivatives being left ones.

Now we are in a position to represent Eq. (65) as a path integral. 
Making use of the Schwinger formula for an inverse operator
\begin{equation}
B^{-1}=i\int\limits_0^\infty ds\,e^{-is(B-i\varepsilon)}, \qquad
\varepsilon\to 0,
\end{equation}
and integration over odd variables (below we use Eq. (67) for the
odd operator $F$, hence $\chi$ and $F$ will anticommute by definition)
\begin{equation}
F=-i\int d \chi e^{i\chi F},
\end{equation}
one gets\footnote{Note that unitarity of the
$G$ requires $\chi$ to be real Grassmann variable while $\mu,\nu$
to be imaginary ones. In what follows
we omit the infinitesimal quantity $\varepsilon$.}
$$
\begin{array}{l}
G(z_{out},z_{in})=\frac 1{32 \sqrt 2} 
e^{\gamma^+ \frac \partial{\partial\mu}} e^{\gamma^-\frac 
\partial{\partial\nu}}\int\limits_0^\infty ds \int d\chi 
\langle z_{out}|e^{-i \hat H(s,\chi)}|z_{in}\rangle|_{\mu=\nu=0},
\end{array}
\eqno{(68a)}$$
$$
\begin{array}{l}
\hat H(s,\chi)=\frac s2 {\hat p}^2-\chi\mu 
{\hat{p'_{\theta}}}^2 {\hat{p'_{\bar\theta}}}^2-\chi\nu      
{\hat{p'_{\bar\theta}}}^2 {\hat{p'_{\theta}}}^2.
\end{array}
\eqno{(68b)}$$
\addtocounter{equation}{1}
After the standard substitution like $e^{-i\hat H}={[e^{-i\hat H/N}]}^N$
and the repeated use of the completeness relation $\int d^8 z |z \rangle 
\langle z|=1$ this yields 
\begin{eqnarray}
&&
G(z_{out},z_{in})=\frac 1{32 \sqrt 2} 
e^{\gamma^+ \frac \partial{\partial\mu}} 
e^{\gamma^-\frac \partial{\partial\nu}}
\int\limits_0^\infty ds \int d\chi 
\int \prod_{k=1}^{N-1}d^8 z_k
\langle z_{out}|
e^{-i \hat H(s,\chi)/N}|z_{N-1}\rangle \cr
&&
\times\dots \langle z_1|e^{-i \hat H(s,\chi)/N}|z_{in}\rangle
|_{\mu=\nu=0}.
\end{eqnarray}
It is worth mentioning that the second term entering in Eq. $(68b)$ is a
$p_\theta \theta$-- and $\bar\theta p_{\bar\theta}$--ordered operator.
Vice versa, the third term is a $\theta p_\theta$-- and
$p_{\bar\theta} \bar\theta$--ordered one. This provides the following
integral representation (see Ref. 36) for the matrix elements involved into 
Eq. (69) (note that the integration over odd variable $\chi$ in 
Eq. (69) guarantees that symbol of the exponential can be replaced by 
the exponential of the symbol)
\begin{eqnarray}
&&
\langle z_k|e^{-i \hat H(s,\chi)/N}|z_{k-1}\rangle=
\frac 1{\pi^4} \int d^8 p_k \exp \{ip_k(z_k-z_{k-1})- 
\frac {is}{2N} {p_k}^2\cr
&&
\qquad +i\frac {\chi\mu}{N} {({p_\theta}_k-
ip_{nk} \sigma^n {\bar\theta}_k)}^2 {(p_{\bar\theta k}+
ip_{nk} \theta_{k-1}\sigma^n)}^2\cr
&&
\qquad \qquad +i\frac {\chi\nu}{N} {(p_{\theta k}-
ip_{nk} \sigma^n {\bar\theta}_{k-1})}^2 
{(p_{\bar\theta k}+ip_{nk} \theta_k \sigma^n)}^2 \}= \cr
&&
=\frac 1{\pi^4} \int d^8 p_k \exp \{ip_k(z_k-z_{k-1})-
\frac {is}{2N} {p_k}^2 
+i\frac {\chi\mu}{N} f_{(\bar l r)k}
+i\frac {\chi\nu}{N} f_{(l\bar r) k} \}, \nonumber \\  
\end{eqnarray}
where we formally attached the labels $l,r$ to the classical function
\linebreak
$f(\theta, p_\theta ; \bar\theta, p_{\bar\theta})\equiv
{(p_\theta-ip_n \sigma^n \bar\theta)}^2 {(p_{\bar\theta}+
ip_n \theta \sigma^n)}^2$ which specify the calculation prescription
for the latter. In particular, $\bar l$ means that 
the argument $\bar\theta$ of the function $f$  
should be taken in the left point of the interval $[z_k,z_{k-1}]$,
while $r$ implies the same for $\theta$ in the right point. In
considering a path integral below, these labels will determine
the descretization prescription.

Now, it only remains to attach indecies to the variables 
$s,\chi,\mu,\nu$. For this purpose we can employ the technique 
developed in Refs. [7,8]. In particular, the ordinary integration over 
$s$ variable can be transformed into the path integral by introducing 
$N$ additional integrations over new auxiliary bosonic variables
$e_k$, $k=1,\dots,N$ followed by the subsequent use of integral
representation for the $\delta$-function
\begin{eqnarray}
&&
G(z_{out},z_{in})=\frac 1{32 \sqrt 2} 
e^{\gamma^+ \frac \partial{\partial\mu}} 
e^{\gamma^-\frac \partial{\partial\nu}}
\int\limits_0^\infty ds \int d\chi 
\int \prod_{k=1}^{N-1}d^8 z_k \prod_{k=1}^{N} 
\frac {d^8 p_k}{\pi^4}
\prod_{k=1}^{N}{de}_k \cr
&&
\times\exp \{ ip_N(z_N-z_{N-1})-i\frac {e_N}{2N} {p_N}^2
+i\frac {\chi\mu}{N} f_{(\bar l r)N}
+i\frac {\chi\nu}{N} f_{(l\bar  r) N} \} \dots \cr
&&
\times\exp \{ ip_1(z_1-z_0)-i\frac {e_1}{2N} {p_1}^2
+i\frac {\chi\mu}{N} f_{(\bar l r)1}
+i\frac {\chi\nu}{N} f_{(l\bar  r) 1} \} \cr
&&
\times\delta(e_N-e_{N-1})\delta(e_{N-1}-e_{N-2})\dots
\delta(e_1-s)|_{\mu=\nu=0}=\cr
&&
=\frac 1{32 \sqrt 2} 
e^{\gamma^+ \frac \partial{\partial\mu}} 
e^{\gamma^-\frac \partial{\partial\nu}}
\int\limits_0^\infty ds \int d\chi 
\int \prod_{k=1}^{N-1}d^8 z_k \prod_{k=1}^{N} 
\frac {d^8 p_k}{\pi^4}
\prod_{k=1}^{N}{de}_k \prod_{k=1}^{N} 
\frac {{d\pi}_k}{2\pi}\cr
&&
\times\exp \{ ip_N(z_N-z_{N-1})
-i\frac {e_N}{2N} {p_N}^2
+i\frac {\chi\mu}{N} f_{(\bar l r)N}
+i\frac {\chi\nu}{N} f_{(l\bar  r) N}
+i\pi_N (e_N-e_{N-1}) \} \cr
&&
\times\dots\exp \{ ip_1(z_1-z_0)-i\frac {e_1}{2N} {p_1}^2
+i\frac {\chi\mu}{N} f_{(\bar l r)1}
+i\frac {\chi\nu}{N} f_{(l\bar  r) 1}+i\pi_1 (e_1-e_0) \}
|_{\mu=\nu=0}, \nonumber \\
\end{eqnarray}
where $z_N\equiv z_{out},z_0\equiv z_{in}, e_0\equiv s$. The odd variables
$\chi,\mu,\nu$ can be treated in a similar way and we finally get
\begin{eqnarray}
&&
G(z_{out},z_{in})=\frac 1{32 \sqrt 2} 
e^{\gamma^+ \frac \partial{\partial\mu}} e^{\gamma^-\frac 
\partial{\partial\nu}}\int\limits_0^\infty ds \int d\chi 
\int \prod_{k=1}^{N-1}d^8 z_k \prod_{k=1}^{N} \frac {d^8 p_k}{\pi^4}
\prod_{k=1}^{N}{de}_k \cr
&&
\times\prod_{k=1}^{N} \frac {{d\pi}_k}{2\pi}
\prod_{k=1}^{N}{(d \lambda_k d \xi_k)} 
\prod_{k=1}^{N}{(d \sigma_k d \omega_k)} 
\prod_{k=1}^{N}{(d {\tilde\sigma}_k d {\tilde\omega}_k)}\cr 
&&
\times\exp i\sum_{k=1}^N \{ p_k(z_k-z_{k-1})/\Delta t-\frac {e_k}{2} {p_k}^2
+\lambda_k \sigma_k f_{(\bar l r)k}
+\lambda_k {\tilde\sigma}_k f_{(l\bar  r) k}\cr
&&
+\pi_k (e_k-e_{k-1})/\Delta t-i\xi_k (\lambda_k-\lambda_{k-1})/\Delta t
-i\omega_k (\sigma_k-\sigma_{k-1})/\Delta t \cr
&&
-i{\tilde\omega}_k({\tilde\sigma}_k-{\tilde\sigma}_{k-1})/\Delta t \}\Delta t\
|_{\mu=\nu=0},
\end{eqnarray}
where $\Delta t\equiv\frac 1N$ and it is implied that the boundary 
conditions 
\begin{eqnarray}
&&
\qquad z_0=z_{in},\qquad z_N=z_{out}, \cr
&&
\qquad e_0=s, \qquad \lambda_0=\chi, \cr
&&
\qquad \sigma_0=\mu, \qquad \tilde\sigma_0=\nu,
\end{eqnarray}
are satisfied.

In the limit of large $N$ one arrives at the path integral
$$
\begin{array}{l}
G(z_{out},z_{in})=\frac 1{32 \sqrt 2} 
e^{\gamma^+ \frac \partial{\partial\mu}} e^{\gamma^-\frac 
\partial{\partial\nu}}\int\limits_0^\infty ds \int d\chi 
\int Dz\,Dp\,De\,D\pi\,D\lambda\,D\xi\,D\sigma\,D\omega\,\times \\
\qquad \qquad \times D\tilde\sigma\,D\tilde\omega\,e^{i(S_{ssp}+S_{gf})},
\end{array}
\eqno{(74a)}$$
$$
\begin{array}{l}
S_{ssp}=\int\limits_0^1 d\tau (p_n\dot x^n+
{p_\theta}_\alpha {\dot\theta}^\alpha+
{p_{\bar\theta}}^{\dot\alpha}{\dot{\bar\theta}}_{\dot\alpha}
-\frac e2 p^2 \\
\qquad +\lambda\sigma {{(p_\theta-ip_n \sigma^n \bar\theta)}^2  
{(p_{\bar\theta}+ip_n \theta \sigma^n)}^2}_{\bar l r} \\
\qquad \qquad +\lambda\tilde\sigma {{(p_\theta-ip_n \sigma^n \bar\theta)}^2 
{(p_{\bar\theta}+ip_n \theta \sigma^n)}^2}_{l \bar r}),
\end{array}
\eqno{(74b)}$$
$$
\begin{array}{l}
S_{gf}=\int\limits_0^1 d\tau (\pi\dot e-i\xi\dot\lambda
-i\omega\dot\sigma-i\tilde\omega\dot{\tilde\sigma}),
\end{array}
\eqno{(74c)}$$
provided the boundary conditions 
$$
\begin{array}{ll}
\qquad \qquad z(0)=z_{in}, \qquad z(1)=z_{out},\\ 
\qquad \qquad e(0)=s, \qquad \lambda(0)=\chi,\\
\qquad  \qquad \sigma(0)=\mu, \qquad \tilde\sigma(0)=\nu,
\end{array}
\eqno{(74d)}$$
hold.
Denoting in Eq. $(74b)$ $p_\theta=\rho$, $p_{\bar\theta}=\bar\rho$ 
we recover the classical theory (18) in the gauge $\psi=0$, $\bar\psi=0$.
The role of the second term $S_{\rm gf}$ in the path integral $(74a)$ is
to fix additional gauge invariance which enters into the problem when 
adding the variables $\lambda,\sigma,\tilde\sigma$ to the original model.

Some comments are in order. First, the gauge fields $\psi,\bar\psi$
can easily be restored in the action $(74b)$ by including the gauge
fixing conditions $\psi=0$, $\bar\psi=0$ into Eq. (72) via the 
$\delta$--function. Second, the labels $l,r$ entering in Eq. $(74b)$, 
although being inessential in classical theory, play an important role
at the quantum level and determine the descretization prescription for
the path integral $(74a)$. Moreover, the change of the variables
${p'}_\theta=p_\theta-ip_n \sigma^n \bar\theta$, 
${p'}_{\bar\theta}=p_{\bar\theta}+ip_n \theta \sigma^n$ in the path
integral $(74a)$, which would bring the action to the original
Siegel one, is problematic in view of the descretization
prescription. This supports, in particular, the advantage of the
geometric formulation [24]. Third, the higher order fermionic constraints,
although being not independent in $d=4$, make a crucial contribution
into the path integral.

\section{Concluding remarks}

Thus, in this paper we have considered operator quantization of
the Siegel superparticle in ${R}^{4|4}$ flat superspace. Quantum
states of the model were proven to be the sum of on-shell chiral and 
antichiral superfields, the corresponding effective field theory being
the massless Wess--Zumino model in the component form. Path integral 
representation for the superfield propagator was constructed and shown
to involve the Siegel action in a gauge fixed form. As a further 
development, it is tempting to compare the result with that of the 
straightforward $BFV$ quantization combined with the scheme [23].
Another open problem is the generalization of the present analysis to
the case of the model coupled to arbitrary external superfields, where the
construction of the path integral representation for the
superpropagator is known to be much more involved.

Due to the relation to superstring theory, the $10d$ case is of prime
interest. The operator quantization presented in this work is rather
specific in four dimensions. We hope, however, that $BFV$ path integral
quantization will proceed along the same lines both in $4d$ and $10d$.
The results on this subject will be presented elsewhere.

\section*{Acknowledgments}

One of the authors (A.V.G.) is grateful to A.A. Deriglazov and
S.J. Gates Jr. for useful discussions. The work of A.V.G has been
supported by INTAS-RFBR Grant No 95-829 and by FAPESP. D.M.G. thanks 
Brasilian foundation CNPq for permanent support.


\begin{thebibliography}{nn}
\bibitem{1} F.A. Berezin and M.S. Marinov, JETP Lett. {\bf 21} (1975) 320;
Ann. Phys. (N.Y.) {\bf 104} (1977) 336.
\bibitem{2} L. Brink, S. Deser, B. Zumino, P. Di Vecchia, and P. Howe,
Phys. Lett. B {\bf 64} (1976) 435;\\ 
L. Brink, P. Di Vecchia, and P. Howe, Nucl. Phys. B {\bf 118} (1977) 76;\\
R. Casalbuoni, Nuovo Cimento A {\bf 33} (1976) 115;\\ 
A. Barducci, R. Casalbuoni, and L. Lusanna, Nuovo Cimento A
{\bf 35} (1976) 377.
\bibitem{3} P.P. Srivastava, Nuovo Cimento Lett. {\bf 19} (1977) 239;\\
A.P. Balachandran, P. Salomonson, B. Skagerstam and J. Winnberg, 
Phys. Rev. {\bf D15}, (1977) 2308;\\               
V.D. Gershun and V.I. Tkach, Pis'ma Zh. Eksper.Teor.Fiz.{\bf 29}
(1979) 320;\\
A. Barducci and L. Lusanna, Nuovo Cimento A {\bf 77} (1983) 39; 
J. Phys. A {\bf 16} (1983) 1993;\\
J. Gomis, M. Novell, and K. Rafanelli, Phys. Rev. D
{\bf 34} (1986) 1072;\\
P.S. Howe, S. Penati, M. Pernici, and P. Townsend,
Phys. Lett. B {\bf 215} (1988) 255;\\
J.W. van Holten, Z. Phys. C. {\bf 41} (1988) 497.
\bibitem{4} D.M. Gitman and I.V.  Tyutin, Class. Quantum Grav.
{\bf 7} (1990) 2131; JETP Lett. {\bf 51} (1990) 214;\\
D.M. Gitman and A.V. Saa, Mod. Phys. A {\bf 8} (1993) 463;\\
D.M. Gitman, A.E. Goncalves, and I.V. Tyutin, Phys. Rev. D
{\bf 50} (1994) 5439; Int. J. Mod. Phys. A {\bf 10} (1995) 701;
Phys. of Atom. Nucl. {\bf 60} (1997) 748.
\bibitem{5}
D.M. Gitman and I.V.  Tyutin, Int. J. Mod.
Phys.  A {\bf 12} (1997) 535; Mod.  Phys.  Lett. A {\bf 11} (1996) 381;\\
D.M. Gitman and A.E. Goncalves, J. Math. Phys.  {\bf 38} 
(1997) 2167; Int. J. Theor. Phys. {\bf 35} (1996) 2427;\\
D.M. Gitman, Nucl. Phys. B (Proc. Suppl.) {\bf 57} (1997) 231;\\
{\it Pseudoclassical theory of relativistic spinning particle}, in Topics
in Satistical and Theoretical Physics, F.A. Berezin Memorial vol.
(Amer. Math. Soc., Providence, PI, 1996). 
\bibitem{6} M.S. Plyushchay, Phys.  Lett. B {\bf 236} (1990) 291;
B {\bf 248} (1990) 299; Mod. Phys.  Lett. A {\bf 8} (1993) 937;\\
R. Marnelius and U. M\"artensson, Nucl. Phys. B {\bf 335}
(1991) 395; Int. J. Mod. Phys. A  {\bf 6} (1991) 807;\\
J.W. van Holten, Int. J. Mod. Phys. A {\bf 7} (1992) 7119;\\
S.M. Kuzenko, S.L. Lyakhovich, and A. Yu. Segal,
Int. J.  Mod. Phys.  A {\bf 10} (1995) 1529;\\
G. Grigorian, R.  Grigorian, and I.V. Tyutin, Teor.  Mat.
Fiz. {\bf 111} (1997) 389.
\bibitem{7} D.M. Gitman, Nucl. Phys. B {\bf 488} (1997) 490.
\bibitem{8} E.S. Fradkin and D.M. Gitman, Phys. Rev.  D. {\bf 44}
(1991) 3230;\\
D.M. Gitman and A.V. Saa, Class. Quan. Grav. {\bf 10} (1993) 1447;\\
D.M. Gitman and Sh.M. Shvartsman, Phys. Lett. B {\bf 318} (1993) 122;\\
D.M. Gitman and S.I. Zlatev, Phys. Rev. D {\bf 55} (1997) 7701.
\bibitem{9}A.T. Ogielski and J. Sobczuk, J. Math. Phys. {\bf 22}
(1981) 2060;\\ 
M. Henneaux, C. Teitelboim, Ann. Phys. {\bf 143} (1982)
127;\\ 
N.V. Borisov and P.P. Kulish, Teor. Math. Fiz. {\bf 51}
(1982) 335;\\
 A.M. Polyakov, {\em Gauge Fields and Strings}, (Harwood,
Chur, Switzerland, 1987);\\
J.C. Henty, P.S. Howe, and P.K. Townsend, Class. Quan. Grav.
{\bf 5} (1988) 807;\\
V.Ya. Fainberg and A.V. Marshakov, JETP
Lett. {\bf 47} (1988) 565;  Phys. Lett. {\bf 211B}  (1988) 81; 
Nucl. Phys. {\bf B306} (1988) 659;\\ 
T.M. Aliev, V.Ya. Fainberg and N.K. Pak, Nucl. Phys. {\bf B429} (1994)
321;\\ 
E. D'Hoker and D.G. Gagn\'e, Nucl. Phys. {\bf B467}
(1996) 272; 297.
\bibitem{10}J.W. van Holten, Nucl. Phys. {\bf B457} (1995) 375;\\
NIKHEF-H/95-055; Proceedings of 29th Int. Symposium on the Theory of 
Elementary Particles, Buckow-1995;\\
{\it Grassmann algebras and spin in quantum dynamics}, Lecture
notes, AIO School Math. Phys.,(Univ. of Twente, 1992).
\bibitem{11} L. Brink, M. Henneaux, and C. Teitelboim, Nucl. Phys. B
{\bf 293} (1987) 505;\\
C.M. Hull and J.L. V\'azquez-Bello, Nucl. Phys. B {\bf 416}
(1994) 173;\\
A.A. Deriglazov, A.V. Galajinsky, and S.L. Lyakhovich, Nucl. Phys. B {\bf
473} (1996) 245.
\bibitem{12} R. Casalbuoni, Phys.  Lett. B {\bf 62} (1976) 49.\\
L. Brink and J.H. Schwarz, Phys. Lett. B {\bf 100} (1981) 310.
\bibitem{13} Y. Eisenberg and S. Solomon, Nucl.  Phys.  B {\bf 309} 
(1988) 709; \\ 
F. Delduc, A. Galperin, and E.  Sokatchev, Nucl. Phys. B {\bf 368} (1992)
143;\\ 
V.P. Akulov, D.P. Sorokin, and I.A. Bandos, Mod. Phys. Lett.  
A {\bf3} (1988) 1633.
\bibitem{14} W. Siegel, Class. Quant. Grav.
{\bf 2} (1985) L 95.
\bibitem{15} T.J. Allen, Mod. Phys. Lett.  A {\bf 2} (1987) 209.
\bibitem{16} W. Siegel, Nucl. Phys. B {\bf 263} (1985) 93; Phys.
Lett. B {\bf 203} (1988) 79.
\bibitem{17} A.R. Mikovi\'{c} and W. Siegel, Phys. Lett. B {\bf 209}
(1988) 47.
\bibitem{18} R. Kallosh, W. Troost, and A.  Van Proeyen, Phys.
Lett. B {\bf 212} (1988) 212.
\bibitem{19} U. Lindstr\"om, M. Rocek, W. Siegel, P. van Nieuwenhuizen, and
A.E. van de Ven, J. Math. Phys. {\bf 31} (1990) 1761;\\
A. Mikovi\'{c}, M. Rocek, W. Siegel, P. van Nieuwenhuizen,
J. Yamron, and A.E. van de Ven, Phys. Lett. B {\bf 235} (1990) 106;\\
F. Ebler, M. Hatsuda, E. Laenen, W.  Siegel, J.P. Yamron, T. Kimura,
and A.  Mikovi\'{c}, Nucl. Phys. B {\bf 364} (1991) 67.
\bibitem{20} M.B. Green and C.M. Hull, Nucl. Phys. B {\bf 344} (1990) 115.
\bibitem{21} A. Duncan and M. Moshe, Nucl. Phys. B {\bf 268} (1986) 706.
\bibitem{22} E.S. Fradkin and Sh.M. Shvartsman, Mod. Phys. Lett. A
{\bf 6} (1991) 1977.
\bibitem{23} A.A. Deriglazov and A.V. Galajinsky, Phys. Lett.
B {\bf 381} (1996) 105; Phys. Rev. D {\bf 54} (1996) 5195.
\bibitem{24} A.A. Deriglazov and A.V. Galajinsky, Mod. Phys.
Lett. A {\bf 9} (1994) 3445.
\bibitem{25} P. van Nieuwenhuizen, Phys. Rep. {\bf 68} (1981) 189.
\bibitem{26} I.L. Buchbinder and S.M. Kuzenko, {\it Ideas and Methods
of Supersymmetry and Supergravity}, (Institute of Physics Publishing,
Bristol and Philadelphia, 1995).
\bibitem{27} P.A.M. Dirac, {\it Lectures on Quantum Mechanics},
Yeshiva University, Belfer Graduate School of Science (Academic Press,
New York, 1964).
\bibitem{28} D.M. Gitman and I.V. Tyutin, {\it Quantization of Fields
with Constraints} (Springer-Verlag, 1990).
\bibitem{29} J. Wess and B. Zumino, Nucl. Phys. B {\bf 70}
(1974) 39.
\bibitem{30} J. Wess and J. Bagger {\it Supersymmetry and Supergravity},
(Princeton University Press, Princeton, 1983).
\bibitem{31} S. Aoyama, J. Kowalski--Glikman, J.  Lukierski, and J.W.
van Holten, Phys.  Lett. B {\bf 216} (1989) 133; B {\bf 201} (1988) 487.
\bibitem{32} S. Bellucci and A.V. Galajinsky, hep-th/9712247, 
Phys. Lett. B (to appear).
\bibitem{33} R. Marnelius and Sh.M. Shvartsman, Nucl.  Phys. B {\bf
430} (1994) 153.
\bibitem{34} I. Bengtsson, Phys. Rev. D {\bf 39} (1989) 1158.
\bibitem{35} J. Schwinger, Phys. Rev. 82 (1951) 664.
\bibitem{36} F.A. Berezin, Uspekhi Fiz. Nauk. {\bf 132} (1980) 497;\\ 
F.A. Berezin and M.A. Shubin, {\em Schr\"odinger Equation} (Moscow State 
University, Moscow, 1983).
\end{thebibliography}
\end{document}